\documentclass[aps,onecolumn,prc,10pt,showpacs,preprintnumbers,
               nofootinbib,float,superscriptaddress,longbibliography]{revtex4-1}
\usepackage{geometry}
\usepackage{epsfig}
\usepackage{color}
\usepackage{float,amsmath,amssymb,diagbox}
\usepackage{multirow}
\usepackage{graphicx}
\usepackage{url}
\usepackage[utf8]{inputenc}
\usepackage{hyperref}
\usepackage{lineno}
\RequirePackage{graphicx}
\usepackage{slashed}
\usepackage{cleveref}

\begin{document}

\title{\LARGE Energy Dependence of Polarized $\mathbf{\gamma\gamma\rightarrow e^{+}e^{-}}$ \newline in Peripheral Au+Au Collisions at RHIC}
\author{The STAR Collaboration}

\date{\today}

\begin{abstract}
We report the differential yields at mid-rapidity of the Breit-Wheeler process ($\gamma\gamma\rightarrow e^{+}e^{-}$) in peripheral Au+Au collisions at $\ensuremath{\sqrt{s_{_{\rm{NN}}}}} = $ 54.4~GeV and 200~GeV with the STAR experiment at RHIC, as a function of energy $\ensuremath{\sqrt{s_{_{\rm{NN}}}}}$, $e^{+}e^{-}$ transverse momentum $p_{\rm T}$, $p_{\rm T}^{2}$, invariant mass $M_{ee}$ and azimuthal angle. In the invariant mass range of 0.4 $<$ $M_{ee}$ $<$~2.6~GeV/$c^{2}$ at low transverse momentum ($p_{\rm T}$ $<$~0.15~GeV/$c$), the yields increase while the pair $\sqrt{\langle p_{\rm T}^{2} \rangle}$ decreases with increasing $\ensuremath{\sqrt{s_{_{\rm{NN}}}}}$, a feature is correctly predicted by the QED calculation. The energy dependencies of the measured quantities are sensitive to the nuclear form factor, infrared divergence and photon polarization. The data are compiled and used to extract the charge radius of the Au nucleus. 
\end{abstract}

\maketitle

\section{Introduction}
\label{intro}
In ultra-relativistic heavy-ion collisions, strong electromagnetic fields arising from the Lorentz contraction of highly charged nuclei generate a large flux of high-energy quasi-real photons (Equivalent Photon Approximation, EPA)~\cite{vonWeizsacker:1934nji,Williams:1934ad}. In collisions of identical nuclei, the photon density is proportional to the square of the ion charge number (Z). Dileptons can be produced via photon-photon interactions ($\gamma\gamma\rightarrow l^{+}l^{-}$) even in ultra-peripheral heavy-ion collisions (UPCs) for which the impact parameter between the colliding nuclei is larger than the sum of their radii such that no nuclear overlap occurs~\cite{Baur:2001jj,Bertulani:2005ru,Baltz:2007kq,Klein:2020fmr}. From the EPA, the photons are preferentially aligned along the collision axis and have transverse momentum on the scale of $\omega/\gamma_{L}$, where $\omega$ is the photon energy and $\gamma_{L}$ is the Lorentz factor of the colliding nuclei. Therefore, the leptons produced by these photon-photon processes have the distinctive signature of being nearly back-to-back in azimuth with small pair transverse momenta~\cite{Zha:2018tlq}. Traditionally these photon-photon fusion processes have been studied only in UPCs~\cite{Vidovic:1992ik,STAR:2004bzo,ALICE:2013wjo,Klein:2016yzr,PhysRevC.80.044902,Hencken:1995me}. However, it has recently been observed that even in hadronic heavy-ion collisions (HHICs), the dilepton production at very low transverse momentum ($p_{\rm T}$) originates mainly from two-photon interactions~\cite{STAR:2018ldd,ATLAS:2018pfw,Lehner:2019amb}. Furthermore, the STAR collaboration at the Relativistic Heavy-ion Collider (RHIC)~\cite{STAR:2018ldd} and the ATLAS collaboration at the Large Hadron Collider (LHC)~\cite{ATLAS:2018pfw} have found a significant pair $p_{\rm T}$ broadening effect for the lepton pairs from photon-photon collisions in HHICs compared to those in UPCs. Previously, it was believed that the transverse momentum distribution of dileptons from the two-photon process should not depend on the impact parameter, and the observed broadening of $p_{\rm T}$ was explained by introducing the final-state effect of either the Lorentz force from a trapped electromagnetic field~\cite{STAR:2018ldd} or Coulomb scattering~\cite{ATLAS:2018pfw} in the Quark-Gluon Plasma (QGP) created in the HHICs. In contrast to these expectations, recent measurements in Pb+Pb UPCs by the CMS collaboration, where the final-state effects are absent, show that the dimuons produced via two-photon process have significant impact parameter dependence~\cite{CMS:2020skx}.
CMS also measured the $p_{\rm T}$ broadening effect, which has quantitatively been described by the generalized EPA (gEPA), lowest order QED, and Wigner function formalism, each of which includes the impact parameter dependence~\cite{Zha:2018tlq,Brandenburg:2021lnj,STAR:2019wlg,Klusek-Gawenda:2020eja,Lin:2022flv}. The broadening effect due to the initial QED field strength should be considered in studying possible trapped magnetic field and multiple scattering in QGP. Specifically, QED calculations with the impact parameter dependence of initial photon kinematics predict a systematically lower $\sqrt{\langle p_{\rm T}^{2}\rangle}$ than the STAR data~\cite{Zha:2018tlq} in HHICs. More experimental studies in the peripheral HHICs are crucial to understand such a discrepancy and to investigate the potential final-state effects.

According to the EPA~\cite{Vidovic:1992ik,Klein:2016yzr}, the photon number density as a function of energy $\omega$ is determined by the field of a single nucleus:
    \begin{equation}
    \label{equation_photon_density}
    n(\omega) = \frac{(Ze)^{2}}{\pi\omega}\int_{0}^{\infty}\frac{d^{2}k_{\perp}}{(2\pi)^{2}}\left[\frac{F\left(\left(\frac{\omega}{\gamma_{L}}\right)^{2}+\overrightarrow{k}_{\perp}^{2}\right)}{\left(\frac{\omega}{\gamma_{L}}\right)^{2}+\overrightarrow{k}_{\perp}^{2}}\right]^{2}\overrightarrow{k}_{\perp}^{2},
    \end{equation}
where $\overrightarrow{k}_{\perp}$ is the photon transverse momentum, and $F\left(\left(\frac{\omega}{\gamma_{L}}\right)^{2}+\overrightarrow{k}_{\perp}^{2}\right)$ is the nuclear electromagnetic form factor.
The photon-photon process is categorized into three possible interactions according to the virtuality of the photons~\cite{STAR:2019wlg}: the collision of two virtual photons (Landau-Lifschitz process~\cite{Landau:1934zj}); the collision of one virtual and one real photon (Bethe-Heitler process~\cite{Bethe:1934za}); and the collision of two real photons (Breit-Wheeler process~\cite{Breit:1934zz}). The transverse momentum of the photons in UPCs is often considered to be related to the virtuality of the photons~\cite{STAR:2004bzo,Baltz:2007kq,Klein:2020jom,Esnault:2021eyg,Zhao:2021ynd}. Therefore, the two-photon process in UPCs is considered to be the Landau-Lifschitz process. However, in many practical calculations, the Breit-Wheeler formalism is applied as a convenient and practical tool~\cite{Budnev:1975poe} ignoring any possible effect from small virtuality. There is no clear consensus on what is considered as the Breit-Wheeler process in UPCs. Constraints on the available phase space for the photons that may participate in the Breit-Wheeler process in heavy-ion collisions have been recently proposed~\cite{Wang:2022ihj}:
    \begin{equation}
    \label{equation_BW_criterion}
    \omega/\gamma_{L}\, \lesssim\, k_{\perp}\, \lesssim\, 1/R\, \ll\, \omega ,    \end{equation}
where $R$ is the charge radius of the colliding nucleus. Due to the STAR kinematic acceptance requirement of single electron (positron) transverse momentum to be greater than 200 MeV/$c$ at midrapidity (with pseudorapidity $|\eta| <$~1), there may not be sufficient phase space for the Breit-Wheeler process as defined in relation ~\ref{equation_BW_criterion} at low beam energies ($\gamma_{L}\lesssim20$) even though the Breit-Wheeler process dominates at top RHIC energy. Distinctive features of the Breit-Wheeler process have been found in recent STAR measurements for Au+Au collisions at 200~GeV~\cite{STAR:2019wlg}. Because real photons with zero mass can not exist in a helicity $J_{Z} = 0$ state, the produced $e^{+}e^{-}$ pair in a collision of two real photons should have a smooth invariant mass ($M_{ee}$) spectra and single electron (positron) momentum preferentially aligned along the collision axis. STAR also confirmed with a pure fourth-order azimuthal angular modulation that the quasi-real photons originating from EPA of Lorentz contraction of electromagnetic fields are linearly polarized~\cite{STAR:2019wlg}. Furthermore, Eq.~\eqref{equation_photon_density} shows an intriguing factor $((\frac{\omega}{\gamma_{L}})^{2}+\overrightarrow{k}_{\perp}^{2})$ inside the form factor ($F$) and in the denominator. 
It constrains the dielectrons produced by real photon-photon processes to have small total transverse momentum. More importantly, it suggests that the total transverse momentum (related to $\overrightarrow{k}_{\perp}$) increases with decreasing beam energy ($\gamma_{L}$) for a given photon energy ($\omega$)~\cite{Wang:2022ihj}.

In this paper, we report the energy and centrality dependence of the polarized $\gamma\gamma\rightarrow e^{+}e^{-}$ process in peripheral Au+Au collisions at $\ensuremath{\sqrt{s_{_{\rm{NN}}}}} = $ 54.4~GeV and 200~GeV. The $e^{+}e^{-}$ yields are presented as a function of pair transverse momentum $p_{\rm T}$, $M_{ee}$, $p_{\rm T}^{2}$ and $\Delta\phi$ – the difference between the azimuthal angles of the sum and difference of the $e^{+}$ and $e^{-}$ momenta. The measurement of $p_{\rm T}^{2}$ can better reflect whether $p_{\rm T}$ has a broadening effect. The yields and $\sqrt{\langle p_{\rm T}^{2} \rangle}$ are also presented as functions of collision energy. Furthermore, we present the measurement of $\langle \cos(4\Delta\phi)\rangle$ as a function of $p_{\rm T}$ predicted for the Breit-Wheeler photon-photon fusion process. Model calculations are compared with the measurements.  The paper is organized as follows. \Cref{sec:experimental setup and data sets} describes the experimental setup and the data sets used in this analysis.~\Cref{sec:analysis technique} explains in detail the analysis techniques, including event and track selection, centrality definition, electron identification, raw signal reconstruction, background subtraction, detector efficiency correction, hadronic cocktail simulation and systematic uncertainties.~\Cref{sec:result and discussion} presents our results on photon-induced dielectron production yields within the STAR detector acceptance and a comparison to theoretical calculations. Our results and conclusions are summarized in~\Cref{sec:conclusions}.

\section{Experimental Setup and Data Sets}
\label{sec:experimental setup and data sets}

This experiment was conducted at the Relativistic Heavy Ion Collider (RHIC)~\cite{Alekseev:2003sk}, and the data were collected by the Solenoidal Tracker at RHIC (STAR) experiment. The major detector subsystems used in this analysis are the Time Projection Chamber (TPC)~\cite{Anderson:2003ur}, the barrel Time-Of-Flight (TOF)~\cite{star:tof}, and a trigger subsystem: the Vertex Position Detectors (VPDs)~\cite{Llope:2014nva}. The TPC is the main tracking detector, used for the measurements of charged particle momenta and for  particle identification (PID) via ionization energy loss per unit length ($dE/dx$). The TOF system consists of the Barrel TOF (BTOF) detector covering the TPC outer cylinder and the VPDs at the forward pseudorapidity regions. Combining the timing information from the VPD and the BTOF detectors, the flight time of the particle can be calculated. The flight time of the particle is further combined with the track length and momentum, both measured by the TPC, to provide charged particle identification. The datasets of Au+Au collisions taken in 2010 and 2011 at $\ensuremath{\sqrt{s_{_{\rm{NN}}}}} = $ 200~GeV, and those taken in 2017 at 54.4~GeV, are used for this analysis. The minimum bias trigger is defined by requiring a coincidence between the signals from the east and west VPDs and collision-vertex cut applied in data taking, in order to select collision events that took place near the center of the detector.

\begin{figure}
    \centering
    \includegraphics[width=0.5\textwidth] {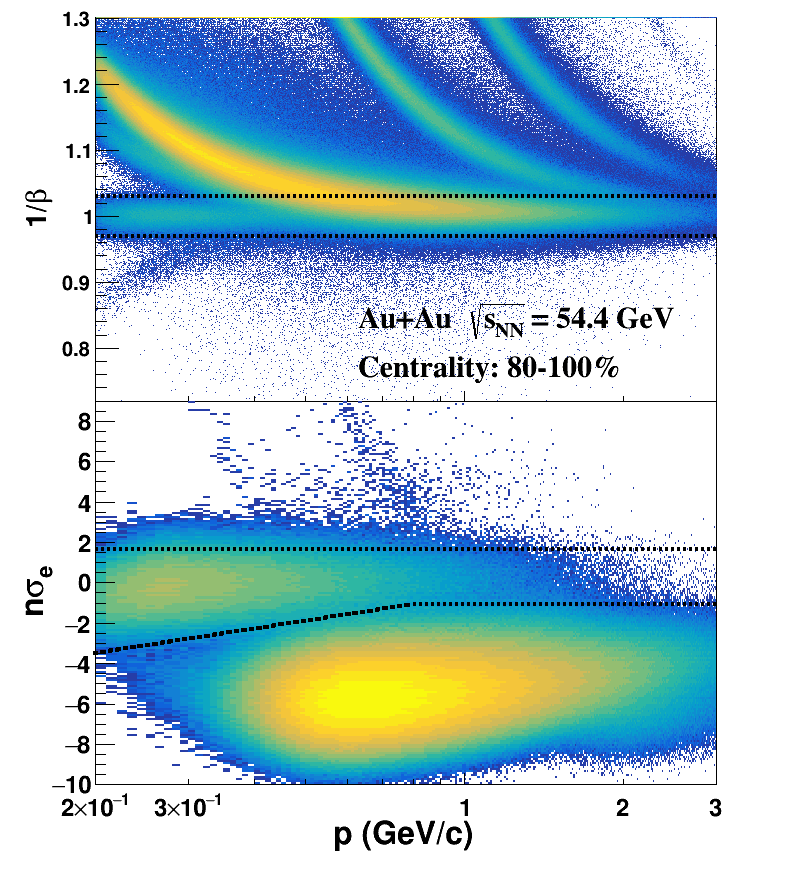}
    \caption{ An example of electron identification at $\ensuremath{\sqrt{s_{_{\rm{NN}}}}} = $ 54.4~GeV in the 80-100\% centrality range. Upper panel: $1/\beta$ vs. momentum ($p$) distributions for all charged particles.  Bottom panel: normalized $dE/dx$ ($n\sigma_{e}$) vs. $p$ distributions after applying the TOF velocity cut $|1 - 1/\beta| <$~0.03 denoted with dashed lines.}
    \label{fig:EID}
\end{figure}

\section{Analysis Techniques}
\label{sec:analysis technique}

\subsection{Event Selection and Centrality Definition}

Events used in this analysis were required to have a reconstructed collision vertex (primary vertex) within 30~cm of the TPC center along the beam direction to ensure uniform detector acceptance, and within a 2~cm radius in the plane perpendicular to the beam direction to reject events hitting the beam pipe (radius: 5~cm). To suppress pile-up events from different bunch crossings in which a TPC vertex was mistakenly reconstructed, the distance along the beam line between the collision vertex constructed using the TPC and that determined by the VPD is required to be less than 3~cm. These selection criteria yield 222~M (year 2010) and 508~M
(year 2011) minimum bias triggered events at $\ensuremath{\sqrt{s_{_{\rm{NN}}}}} = $ 200~GeV and 490~M (year 2017) minimum bias triggered events at $\ensuremath{\sqrt{s_{_{\rm{NN}}}}} = $ 54.4~GeV. The results at $\ensuremath{\sqrt{s_{_{\rm{NN}}}}} = $ 200~GeV reported in this paper are from the combined year 2010 and year 2011 data.

The centrality definition used in this analysis is determined by matching the TPC measured (uncorrected) charged particle multiplicity density $dN/d\eta$ within $|\eta| <$~0.5 with a Monte Carlo Glauber simulation~\cite{Miller:2007ri,STAR:2022tfp}. The centrality bins are defined according to the Monte Carlo Glauber distribution. In particular, if the charged particle multiplicity is less than that corresponding to 80\% centrality, it is defined as 80-100\%.

\subsection{Track Selection}

The main detector subsystems used to reconstruct the electron candidate tracks (including positrons if not specified) are the TPC and TOF. The number of fit points in the TPC (nHitsFit) is required to be at least 15-20 (this cut is different for different runs) to ensure sufficient momentum resolution, and no fewer than 10-16 (this cut is different for different runs) space points (nHitsdEdx) are required for the ionization energy loss ($dE/dx$) calculation to ensure good $dE/dx$ resolution. The ratio of the number of fit points over the number of possible points should be greater than 0.52 in order to avoid track splitting in the TPC. The distance of closest approach to the primary vertex (DCA) is required to be less than 1 cm to reduce the contributions from secondary decays. Each track's transverse momentum should be greater than 0.2~GeV/$c$ to ensure that the track can pass through the TPC. Furthermore, the tracks are required to match to a hit in TOF, which only covers $\sim$90\% in azimuth.

\subsection{Electron Identification}

Electrons were identified by combining the normalized $dE/dx$ from the TPC and velocity ($\beta$) from the TOF. The normalized $dE/dx$ is defined as follows:
    \begin{equation}
    \label{equation_nsigmaE}
    n\sigma_{e} = \frac{\ln(\langle dE/dx \rangle ^{m}/\langle dE/dx \rangle ^{th}_{e})}{R_{dE/dx}} ,
    \end{equation}
where $\langle dE/dx \rangle ^{m}$ and $\langle dE/dx \rangle ^{th}$ represent measured and theoretical $dE/dx$ values, respectively, and $R_{dE/dx}$ is the experimental $dE/dx$ resolution. More details about the electron identification procedure can be found in Refs.~\cite{Shao:2005iu, STAR:2015tnn}. 

The inverse particle velocity ($1/\beta$) measured by the TOF versus the particle momentum $p$ measured by the TPC is shown in the upper panel of Fig.~\ref{fig:EID} for all charged particles in Au+Au collisions at $\ensuremath{\sqrt{s_{_{\rm{NN}}}}} = $ 54.4~GeV in the 80-100\% centrality range. The area enclosed by the two black lines is the TOF velocity selection condition $|1 - 1/\beta| <$~0.03. The bottom panel shows the $n\sigma_{e}$ vs. $p$ distribution after applying the TOF velocity selection. The area enclosed by the black lines is the selection condition from $n\sigma_{e}$. By making use of the measured $n\sigma_{e}$ and $1/\beta$, the electron sample can be selected at a high purity. The electron purity for $p_{\rm T}^{e} >$~0.2~GeV/$c$ is about 95\% in both the $\ensuremath{\sqrt{s_{_{\rm{NN}}}}} = $ 54.4~GeV and $\ensuremath{\sqrt{s_{_{\rm{NN}}}}} = $ 200~GeV data samples.

\subsection{ $e^{+}e^{-}$ Pair Reconstruction and Background Subtraction}
\begin{figure}
    \centering
    \includegraphics[width=0.85\textwidth]{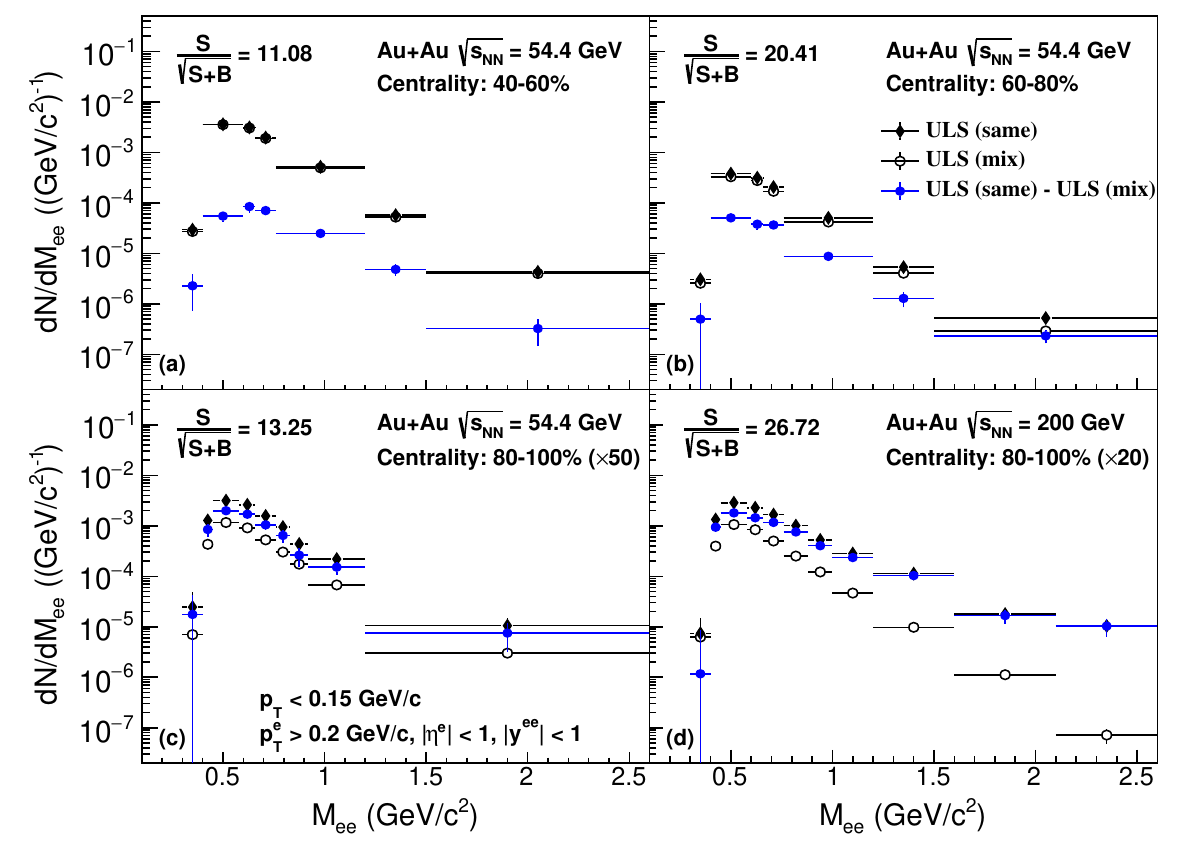}
    \caption{ The low-$p_{\rm T}$ ($p_{\rm T} < $~0.15~GeV/$c$) $e^{+}e^{-}$ raw mass spectra within the STAR acceptance in Au+Au collisions at $\ensuremath{\sqrt{s_{_{\rm{NN}}}}} = $ 54.4~GeV in the centralities (a) 40-60\%, (b) 60-80\%, (c) 80-100\% (scaled $\times$50) and (d) $\ensuremath{\sqrt{s_{_{\rm{NN}}}}} = $ 200~GeV for the 80-100\% centrality range (scaled $\times$20). ULS (same) and ULS (mix) correspond to unlike sign in same event and mixed event, respectively.  Statistical uncertainties are shown as vertical bars.}
    \label{fig:raw mass}
\end{figure}

For each event, all electron and positron candidates within the STAR acceptance of $p_{\rm T}^{e} >$~0.2~GeV/$c$ and $|\eta| <$~1 are combined to generate the (same event) inclusive unlike-sign pairs ($N_{+-}$, including signal and background). In this analysis the signal is defined as the $e^{+}e^{-}$ pairs that originate from photon-photon processes. Background sources that contribute to the inclusive unlike-sign pair distributions include: 

$\bullet$ Combinatorial background pairs, which come from uncorrelated electron and positron pairing. 

$\bullet$ Photon conversion background pairs, which come from photons interacting with the detector material and converting into $e^{+}e^{-}$ pairs.

$\bullet$ Hadronic cocktail background pairs, which originate from hadron decays such as $\pi^{0}$, $\eta$, $\eta^{\prime}$, $\omega$, $\phi$, $J/\psi$, as well as correlated charmed hadrons.

Contributions from combinatorial background pairs are calculated by using the mixed-event unlike-sign pairs. The photon conversion electron pairs are removed from the same event and mixed event using the $\phi_{V}$ cut method~\cite{PHENIX:2009gyd,STAR:2013pwb}. This method relies on the kinematics of the pair production process. The opening angle, $\phi_V$, for electron-positron pairs due to photon conversions should be zero. Unit-vector definitions used for the construction of the $\phi_{V}$ angle were taken from Refs.~\cite{PHENIX:2009gyd,STAR:2015tnn}. Finally, the raw dielectron signal can be obtained by subtracting the mixed-event unlike-sign pairs from the same event unlike-sign pairs. Fig.~\ref{fig:raw mass} shows the low-$p_{\rm T}$ invariant mass distributions of same event unlike-sign pairs (black dots), mixed-event unlike-sign pairs (open circles) and raw dielectron signal (blue dots) for $\ensuremath{\sqrt{s_{_{\rm{NN}}}}} = $ 54.4~GeV and $\ensuremath{\sqrt{s_{_{\rm{NN}}}}} = $ 200~GeV in different centralities. The significance $\frac{S}{\sqrt{S+B}}$ of the signal is also shown in Fig.~\ref{fig:raw mass}, where $S$ and $B$ represent the number of raw signal and background events, respectively.
    
\subsection{Efficiency Correction}    
    
The raw $e^{+}e^{-}$ signal is corrected for the detector efficiency to obtain the final physics $e^{+}e^{-}$ signal. The pair efficiency within STAR acceptance (single electron transverse momentum $p_{\rm T}^{e} >$~0.2~GeV/$c$, single electron pseudorapidity $|\eta_{e}| <$~1, dielectron rapidity $|y^{ee}| < $1) is evaluated from the single electron efficiency by using a Monte Carlo (MC) simulation that used the virtual photons as the input and let them decay into dielectrons isotropically. The single electron efficiency losses are caused by the detector inefficiency and electron identification cuts. 

The detector efficiency includes the TPC tracking efficiency and TOF matching efficiency. The TPC tracking efficiency is evaluated via the standard STAR embedding technique~\cite{STAR:2008med}. The real data electrons from $\pi^{0}$ Dalitz decays and photon conversion are identified by invariant mass and used as the high purity electron sample to evaluate the TOF matching efficiency. Due to the limited statistics of this high purity electron sample, the pure pion sample selected by a tight $n\sigma_{\pi}$ cut is used to generate the three dimensional ($p_{\rm T}$, $\eta$, and $\phi$) TOF matching efficiency. The $p_{\rm T}$-dependent correction factor is then used to correct the TOF matching efficiency difference between electrons and pions caused by the decay loss of pions between the TPC and TOF, as well as other effects. The correction factor is the TOF matching efficiency ratio of electrons to pions as a function of $p_{\rm T}$.

The electron identification efficiency includes two components: TOF $1/\beta$ cut efficiency and $n\sigma_{e}$ selection criteria efficiency. Both of these efficiencies are evaluated using the high purity electron samples identified by invariant mass.

The MC simulation, with virtual photons as input, is used to fold the single electron efficiency into the $e^{+}e^{-}$ pair efficiency within STAR acceptance. The two dimensional kinematics ($M_{ee}$, $p_{\rm T}$) of the virtual photon is taken from the hadronic cocktail (discussed in~\Cref{sec:hadronic cocktail}). The virtual photons have flat rapidity and azimuthal distributions, and decay into $e^{+}e^{-}$ pairs isotropically. The virtual photon simulation is also used to estimate the pair $\phi_{V}$ cut efficiency.

\begin{figure}
    \centering
    \includegraphics[width=0.99\textwidth]{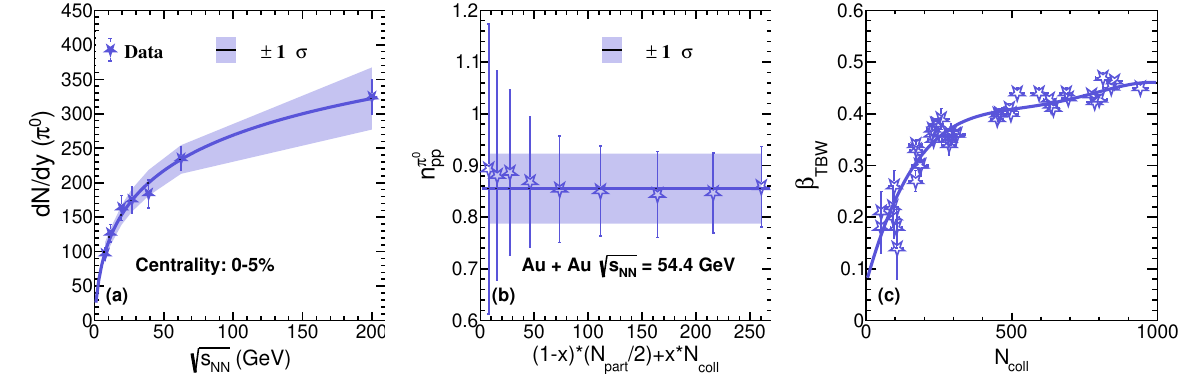}
    \caption{ (a) The $\pi^{0}$ yield~\cite{STAR:2017sal,STAR:2008med} as a function of beam energy in 0-5\% centrality with the exponential fit shown as the solid line. The band is the 1$\sigma$ confidence interval of the fit. (b) The $n_{pp}^{\pi^{0}}$ value in different centralities at 54.4~GeV fitted by a constant shown as the solid line. The band is the 1$\sigma$ confidence interval of the fit. (c) The parameter $\beta_{\rm TBW}$ for Tsallis Blast Wave fits as a function of $N_{\text{coll}}$ with a polynomial fit shown as the solid line.}
\label{fig: pi0 yield and hadron pt}
\end{figure}

\begin{figure}
    \centering
    \includegraphics[width=0.99\textwidth]{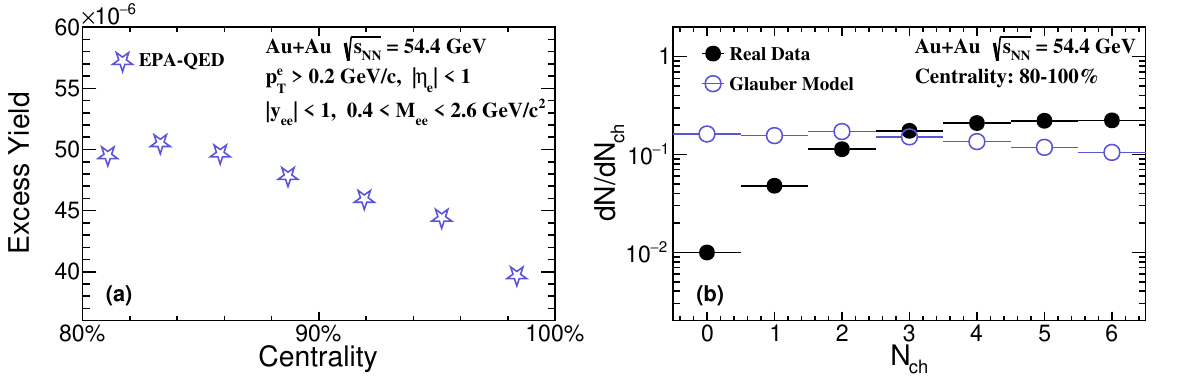}
\caption{ (a) The dielectron yield of the $\gamma\gamma\rightarrow e^{+}e^{-}$ process as a function of centrality at 54.4~GeV calculated by EPA-QED. (b) The multiplicity distribution of charged particles ($\rm N_{ch}$) at 54.4~GeV in 80-100\% centrality for data and a Glauber Model simulation.}
\label{fig: trigger bias}
\end{figure}

\subsection{Hadronic Cocktail}
\label{sec:hadronic cocktail}

The detected $e^{+}e^{-}$ pairs, originating from all stages in the evolution of heavy-ion collisions, are contributed by hadron decays (known as hadronic cocktail) and photon-photon processes. The latter are mainly concentrated at low-$p_{\rm T}$~\cite{STAR:2018ldd}. The hadronic cocktail contributions in the final dielectron spectrum can be evaluated as long as the hadron yields $dN/dy$ and $p_{\rm T}$ distributions are known. The hadronic cocktails included in our simulation contain contributions from decays of $\pi^{0}$, $\eta$, $\eta^{\prime}$, $\omega$, $\phi$, and $J/\psi$, as well as correlated charmed hadrons. The input rapidity distributions are assumed to be flat within $|y| <$~1.2. The input $dN/dy$ and $p_{\rm T}$ distributions are discussed below.

The input $\pi^{0}$ yield is taken as the average of the $\pi^{+}$ and $\pi^{-}$ yields. Other hadron yields are obtained by scaling by the ratio of the hadron to $\pi^{0}$ cross sections~\cite{CERES:2005uih}. There were no measurements of hadron yields at 54.4~GeV, while the $\pi^{\pm}$ yields have been accurately measured in the STAR BES-I for 7.7, 11.5 19.6, 27, and 39~GeV~\cite{STAR:2017sal} and in Ref.~\cite{STAR:2008med} for 62.4 and 200~GeV. The exponential function $p_0*\exp[-(x/p_1)^{p_2}]$ was used to fit the $\pi^{0}$ yields for these energies to interpolate the $\pi^{0}$ yields for the corresponding centrality at 54.4~GeV, where $p_0$, $p_1$ and $p_2$ are parameters.  Fig.~\ref{fig: pi0 yield and hadron pt}a is an example of exponential fitting to interpolate the $\pi^{0}$ yield at 54.4~GeV in 0-5\% centrality; the band is the 1$\sigma$ confidence interval of the fit and was used as a systematic uncertainty in the cocktail. The ``two components model"~\cite{Kharzeev:2000ph} given as:

\begin{equation}\label{eq:two component model}
\frac{dN_{\pi^{0}}}{dy} = n_{pp}^{\pi^{0}}*[(1-x)*\frac{N_{\text{part}}}{2} + x*N_{\text{coll}}],
\end{equation}

\noindent is used to described the particles yield, where $N_{\text{coll}}$ is the number of binary collisions and $N_{\text{part}}$ is the number of participating nucleons. $N_{\text{coll}}$ and $N_{\text{part}}$ are obtained from the Glauber Model~\cite{Miller:2007ri} in our analysis. Therefore, according to the ``two components model"\cite{Kharzeev:2000ph}, $n_{pp}^{\pi^{0}}$ in each centrality at 54.4~GeV can be obtained as shown in Fig.~\ref{fig: pi0 yield and hadron pt}. At a given energy, $n_{pp}^{\pi^{0}}$ should be a constant, so fitting the data in Fig.~\ref{fig: pi0 yield and hadron pt}b with a constant resulted in $n_{pp}^{\pi^{0}} = 0.855 \pm 0.068$ at 54.4~GeV. As before, the band is the 1$\sigma$ confidence interval of the fit and was used as a systematic uncertainty in the cocktail. Based on Eq.~\eqref{eq:two component model}, the yields of $\pi^{0}$ in the centralities 40-60\%, 60-80\% and 80-100\% at 54.4~GeV are 30.81 $\pm$ 2.44, 9.45 $\pm$ 0.75 and 2.74 $\pm$ 0.22, respectively. The charged pion yields for 200~GeV Au+Au minimum-bias collisions have been accurately measured in the STAR acceptance~\cite{STAR:2008med}, so again the ``two component model" was used to extrapolate the $\pi^{0}$ yield in 80-100\% centrality at 200~GeV, where the $\pi^{0}$ yield was found to be 4.32 $\pm$ 0.31.

For light hadrons, the Tsallis Blast-Wave (TBW) functions provide good parameterizations of their $p_{\rm T}$ spectra~\cite{Chen:2020zuw}. Polynomial functions were used to fit the TBW parameters as a function of $N_{\text{coll}}$ to extrapolate the parameters in our study. Fig.~\ref{fig: pi0 yield and hadron pt}c is an example of polynomial fitting to parameter $\beta_{\rm TBW}$ for the flow velocity in TBW. The extracted TBW parameters from the fit are used to generate the light hadron spectra as our input. The cocktail input of the $J/\psi$ $p_{\rm T}$ spectra is the same as that in Ref.~\cite{STAR:2018xaj} at 62.4~GeV. We note that the hadronic cocktail for $J/\psi$ are in a different mass range and phase space, though this should have little effect on our results.

The correlated open-charm decay contributions in $p+p$ collisions were obtained from PYTHIA simulations~\cite{Sjostrand:2000wi} and scaled by $N_{\text{coll}}$ in Au+Au collisions for the default cocktail simulations. It should be noted that the Glauber Model cannot describe experimental data due to trigger bias in peripheral collisions, so weights are included to correct charged particle multiplicities from the data to the Glauber Model. In particular, the experimental data in 80-100\% centrality have not been corrected for trigger bias, and therefore needs to be studied. Fig.~\ref{fig: trigger bias}a shows the dielectron yield as a function of centrality calculated by EPA-QED for the $\gamma \gamma \rightarrow e^{+}e^{-}$ process at 54.4~GeV~\cite{Wang:2022ihj}. Fig.~\ref{fig: trigger bias}b is the multiplicity distribution of charged particles ($\rm N_{ch}$) from our measurement compared to a Glauber model simulation at 54.4~GeV in 80-100\% centrality. By taking the difference of the two distributions in Fig.~\ref{fig: trigger bias}b, the effects of trigger bias on the dielectron yield from the $\gamma \gamma \rightarrow e^{+}e^{-}$ process at 54.4~GeV in 80-100\% centrality can be determined.Thus, we conclude that the definition of 80-100\% centrality at 54.4~GeV has a bias of 4.5\% for the $\gamma \gamma \rightarrow e^{+}e^{-}$ process. Similarly, the definition of 80-100\% centrality at 200~GeV has a bias of 1.3\% for this process.

\begin{figure}
    \centering
    \includegraphics[width=0.85\textwidth]{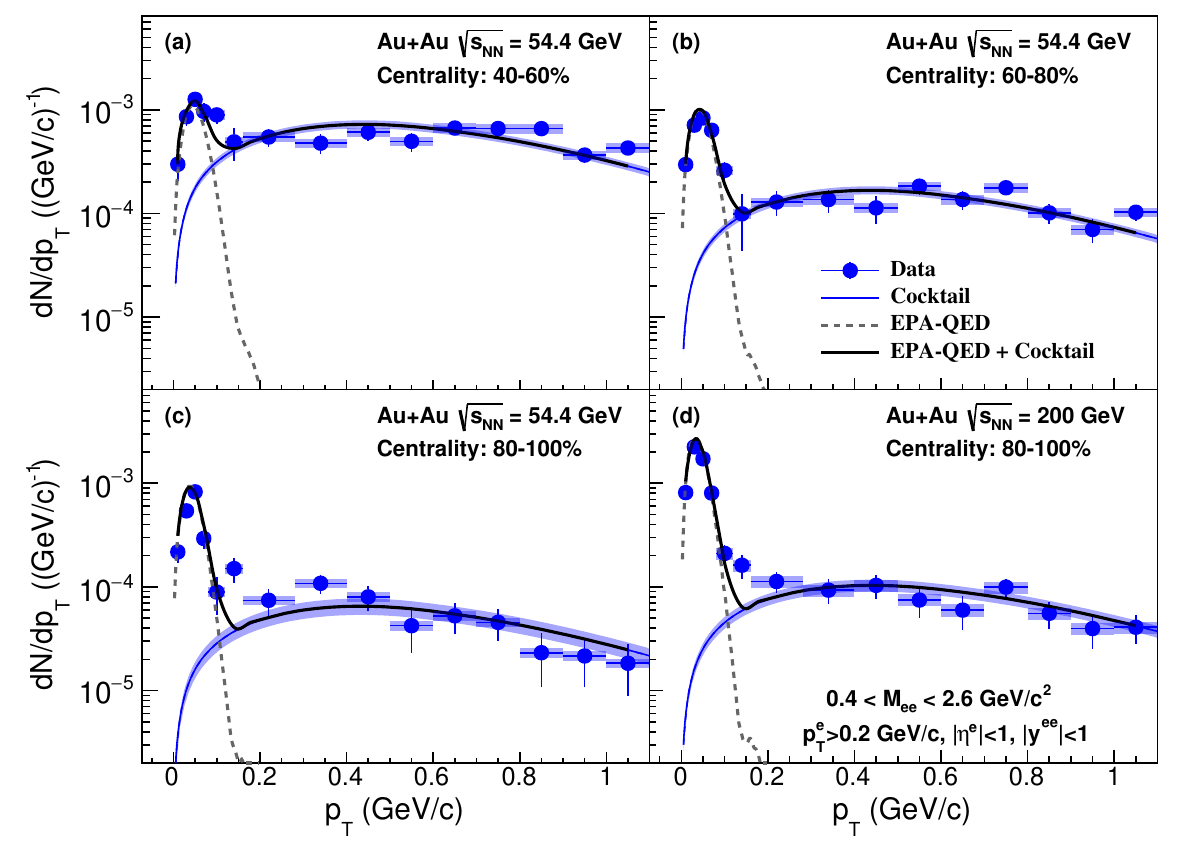}
    \caption{ The $p_{\rm T}$ distributions of $e^{+}e^{-}$ pairs within the STAR acceptance for invariant mass region 0.4-2.6~GeV/$c^{2}$ in Au+Au collisions at $\ensuremath{\sqrt{s_{_{\rm{NN}}}}} = $ 54.4~GeV in the centralities (a) 40-60\%, (b) 60-80\%, (c) 80-100\% and (d) $\ensuremath{\sqrt{s_{_{\rm{NN}}}}} = $ 200~GeV for the 80-100\% centrality range, compared to cocktail (solid blue line) and the lowest order EPA-QED predictions (dashed line)~\cite{Wang:2022ihj}. Statistical uncertainties are shown as vertical bars. The systematic uncertainties of the data are shown as blue boxes. The blue bands depict the systematic uncertainties of the cocktails.}
    \label{fig:pt}
\end{figure}

\begin{table*}[htbp]
  \centering
  \caption{\label{table: Systematic uncertainties on pair efficiency at 54.4 GeV}Systematic uncertainties on pair efficiency and hadron contamination at 54.4~GeV.}
  \setlength{\tabcolsep}{5mm}{
\begin{tabular}{ccc}
 \hline \hline
  & \textbf{Source} & \textbf{Uncertainty (\%)} \\ \hline
 \multirow{4}{*}{\textbf{TPC}} & nHitsFit       & 3.4  \\
                               & nHitsdEdx      & 0.9  \\ 
                               & DCA            & 9.1  \\ 
                               & n$\sigma_{e}$  & 1.1  \\   \hline
 \multirow{2}{*}{\textbf{TOF}} & matching       & 0.1  \\
                               & $1/\beta$      & 3.5  \\   \hline
 \textbf{Hadron Contamination} &                & 2    \\ \hline 
 \textbf{Total}                &                & 10.7 \\ \hline \hline
\end{tabular}}
\end{table*}

\subsection{Systematic Uncertainties}
\label{sec:systematic uncertainties}

The sources of systematic uncertainty that contribute to the final result in this analysis includes: an efficiency correction, hadron contamination, and subtraction of the cocktail contribution. The systematic uncertainty on the efficiency correction comes from the uncertainty on the single-track reconstruction efficiency, which is estimated by comparing the embedding and data. A virtual photon simulation method was used to fold the uncertainty on the single-track efficiency to the pair systematic uncertainty. The pair systematic uncertainties for each individual component of the efficiency correction at $\ensuremath{\sqrt{s_{_{\rm{NN}}}}} = $~54.4~GeV are summarized in \Cref{table: Systematic uncertainties on pair efficiency at 54.4 GeV}. The systematic uncertainty of DCA is relatively large because the embedding does not describe the DCA distribution of the data well at 54.4~GeV. The systematic uncertainty on the efficiency correction at $\ensuremath{\sqrt{s_{_{\rm{NN}}}}} =$ 200~GeV follows  Ref.~\cite{STAR:2015tnn}. The electron candidates contain a small amount of hadron contamination, which may be combined with each other or electrons and thus contribute to the signal pairs. This contribution can be estimated by using the pure pion, kaon and proton samples, which results in an uncertainty of less than 2\% at $\ensuremath{\sqrt{s_{_{\rm{NN}}}}} = $ 54.4~GeV as listed in \Cref{table: Systematic uncertainties on pair efficiency at 54.4 GeV}. The systematic uncertainty from hadron contamination is 5\% at $\ensuremath{\sqrt{s_{_{\rm{NN}}}}} = $ 200~GeV. The total systematic uncertainty of the signal is determined via the quadratic sum of the efficiency correction and hadron contamination. The systematic uncertainties on the cocktail are dominated by the extrapolated uncertainties on particle yields (discussed in~\Cref{sec:hadronic cocktail}) and the uncertainties on the decay branching ratios of hadrons to dielectrons as determined by the Particle Data Group~\cite{ParticleDataGroup:2020ssz}. Another important contribution is that of thermal radiation and $\rho$ meson decays, which cannot be estimated by simulation. It is worth noting that the expected contribution from thermal radiation is also larger toward more central collisions. Given that hadrons freeze out at a particular temperature, and the average temperature from thermal radiation is similar, one expects therefore that the dielectrons from thermal radiation and $\rho$ meson decays should have similar $p_{\rm T}$ distributions. In fact, in-medium $\rho^0$ decay dominates the low-$p_T$ and low-mass dielectron range~\cite{Rapp:2016xzw,STAR:2018xaj}. To take this into account, we scale the cocktail to match the data for $p_{\rm T}$ $\sim$ (0.2, 1.1)~GeV/$c$ and take this extra scaling factor as a systematic uncertainty in the cocktail. These contributions are included in the total systematic uncertainty of signal $e^+e^-$ yield. The scale factors are determined to be 1.33, 1.06, and 1.47 at 40-60\%, 60-80\% and 80-100\% centralities at 54.4~GeV, respectively, and 1.23 in 80-100\% centrality at 200~GeV.

\section{Results and Discussion}
\label{sec:result and discussion}

\begin{figure}
    \centering
    \includegraphics[width=0.85\textwidth]{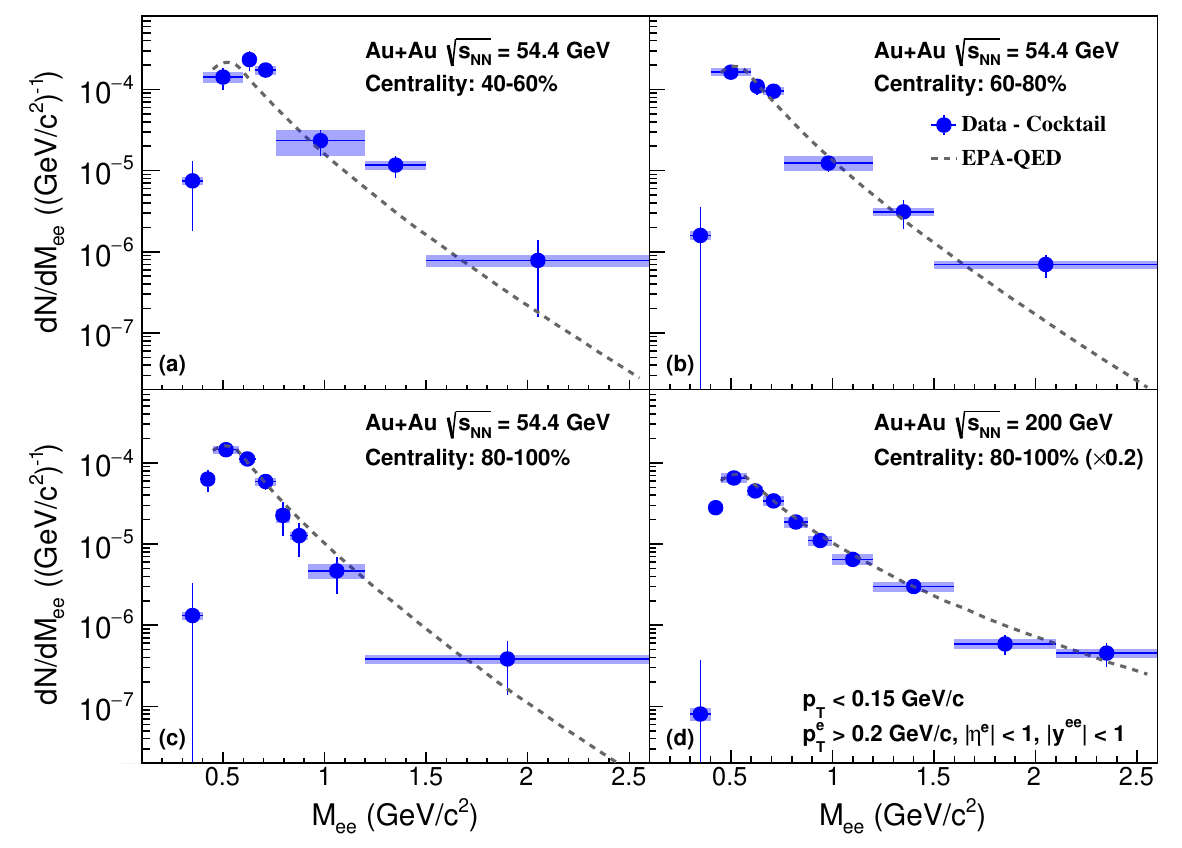}
    \caption{ The low-$p_{\rm T}$ ($p_{\rm T} <$~0.15~GeV/$c$) $e^{+}e^{-}$ excess mass spectra (Data$-$Cocktail) within the STAR acceptance in Au+Au collisions at $\ensuremath{\sqrt{s_{_{\rm{NN}}}}} = $ 54.4~GeV in the centralities (a) 40-60\%, (b) 60-80\%, (c) 80-100\% and (d) $\ensuremath{\sqrt{s_{_{\rm{NN}}}}} = $ 200~GeV for the 80-100\% centrality range (scaled $\times$0.2), compared to the lowest order EPA-QED predictions (dashed line)~\cite{Wang:2022ihj}. Statistical uncertainties are shown as vertical bars, while systematic uncertainties are shown as blue boxes. }
    \label{fig:invariant mass}
\end{figure}

\subsection{Transverse Momentum Distributions}

The $p_{\rm T}$ distributions of $e^{+}e^{-}$ pairs within STAR acceptance ($p_{\rm T}^{e} >$~0.2~GeV/$c$, $|\eta^{e}| <$~1, and $|y^{ee}| <$~1) for the invariant mass region 0.4-2.6~GeV/$c^{2}$ in Au+Au collisions at $\ensuremath{\sqrt{s_{_{\rm{NN}}}}} = $ 54.4~GeV and 200~GeV in different centralities are shown in Fig.~\ref{fig:pt}. A significant enhancement in the yield is found below $p_{\rm T} \approx 0.15$~GeV/$c$ at these energies and centralities, while the hadronic cocktail, shown as the blue curve in Fig.~\ref{fig:pt}, can describe the data reasonably well for $p_{\rm T} >$~0.15~GeV/$c$. These excesses are consistent with the lowest order EPA-QED predictions for the collision of linearly polarized photons quantized from the extremely strong electromagnetic fields generated by the highly charged Au nuclei at ultra-relativistic speed~\cite{Zha:2018tlq}. We note that Fig.~\ref{fig:pt} panels (c) and (d) show that there may be potentially an enhanced yield at $p_T\simeq0.2$~GeV/$c$. We have examined single track and pair azimuthal and $p_T$ distributions for those pairs to see if there is any correlation due to detector defects. We also checked pion contamination in the electron identification. None of these studies indicate that detector effects would cause such enhancement. Future high-statistic datasets with much smaller trigger bias in peripheral Au+Au collisions at 200~GeV may help determine if this is a real physical effect.

\subsection{Invariant Mass Distributions}

\begin{figure}
    \centering
    \includegraphics[width=1.\textwidth]{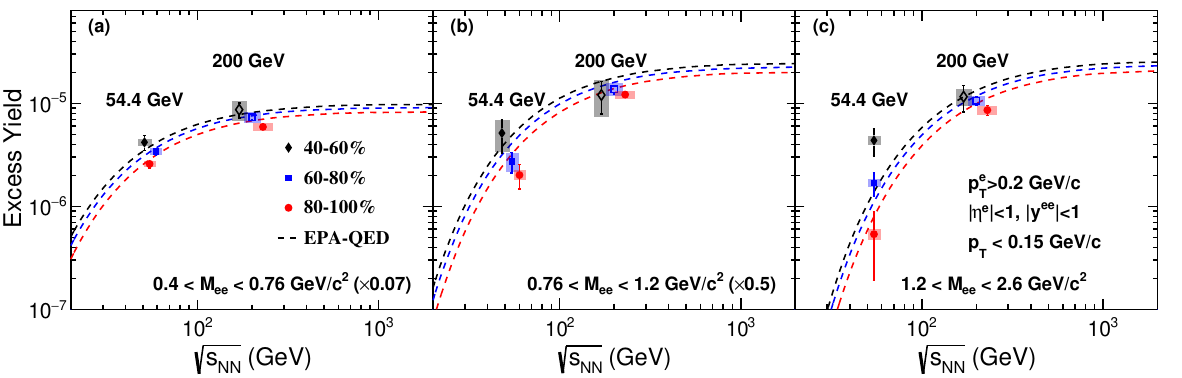}
    \caption{ The integrated signal $e^+e^-$ yields as a function of collision energy in the mass regions of (a) 0.4–0.76~GeV/$c^{2}$  (scaled $\times$0.07), (b) 0.76–1.2~GeV/$c^{2}$  (scaled $\times$0.5), and (c) 1.2–2.6~GeV/$c^{2}$ in Au+Au collisions in the 40-60\%, 60-80\%, and 80-100\% centrality ranges. The points are offset slightly in the horizontal direction for clarity. The energy dependence of integrated signal $e^+e^-$ yields from the lowest order EPA-QED predictions~\cite{Wang:2022ihj} are also shown as dashed lines for comparison. Statistical uncertainties are shown as vertical bars. The systematic uncertainties of the data are shown as boxes. Open markers are extracted from STAR previous published data in Refs.~\cite{STAR:2018ldd}.}
    \label{fig: excess yield}
\end{figure}

After subtracting the hadronic cocktail contribution from the inclusive $e^{+}e^{-}$ pairs, the invariant mass distributions of excess pairs for $p_{\rm T} <$~0.15~GeV/$c$ are shown in Fig.~\ref{fig:invariant mass} for $\ensuremath{\sqrt{s_{_{\rm{NN}}}}} = $ 54.4~GeV and 200~GeV in different centralities. The invariant mass spectra are smooth and featureless even in the range of known vector mesons. This a consequence of the charge parity conservation allowing for $J^{PC}$ states of two photons with positive C-parity only, irrespective of their virtuality. Thus, with the hadronic cocktail contributions removed, the remaining pairs are predominantly due to the photon-photon process of interest.  These signal pairs are also consistent with the lowest order EPA-QED predictions~\cite{Wang:2022ihj}.

We integrated the low-$p_{\rm T}$ invariant mass distributions for signal pairs in the invariant mass regions of (a) 0.4–0.76, (b) 0.76–1.2, and (c) 1.2–2.6~GeV/$c^{2}$. The integrated signal yields as a function of beam energy for the centrality intervals of 40-60\%, 60-80\% and 80-100\% are shown in Fig.~\ref{fig: excess yield}. We note that the signal yields in a given centrality increase with beam energy in all three mass regions. EPA-QED~\cite{Wang:2022ihj} predicts similar energy dependences, which are consistent with the data.

\begin{figure}
    \centering
    \includegraphics[width=0.8\textwidth]{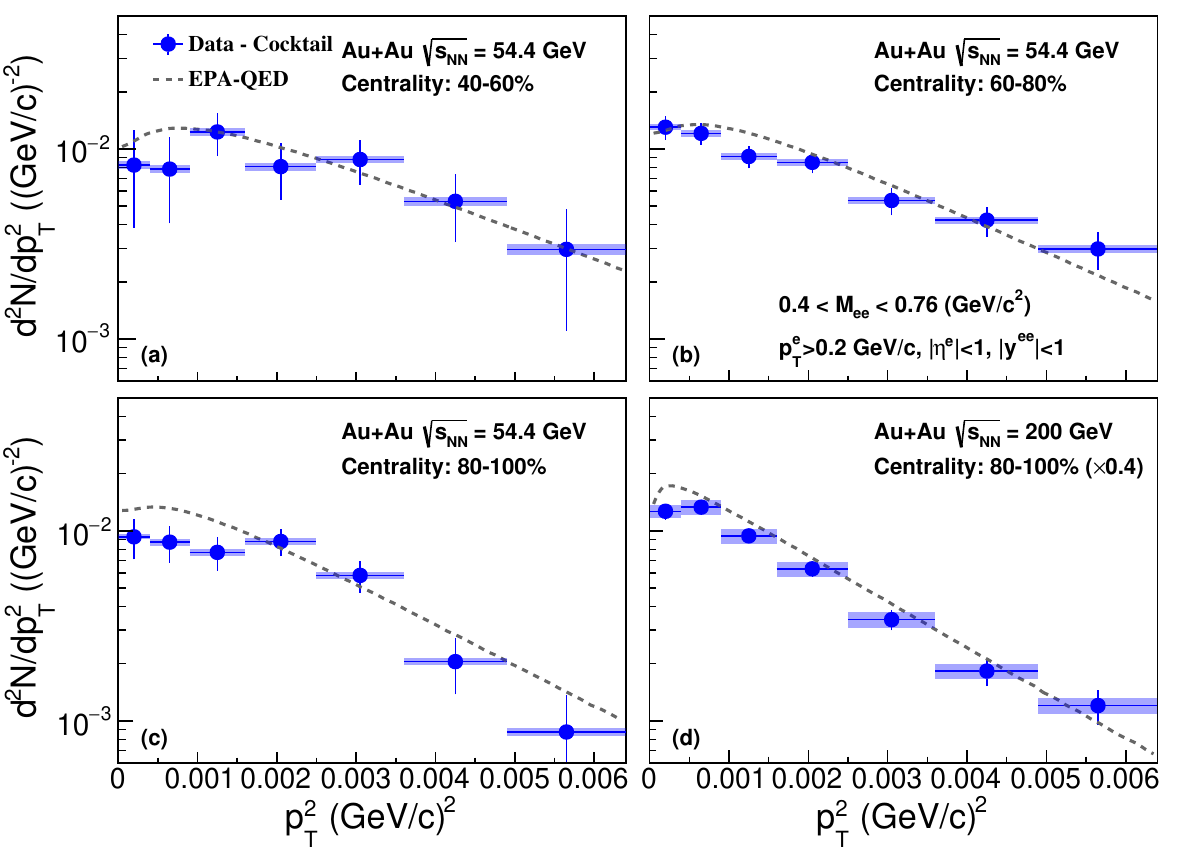}
    \caption{ The $p_{\rm T}^{2}$ distributions of the signal dielectrons within STAR acceptance in the mass region of 0.4-0.76~GeV/$c^{2}$ at $\ensuremath{\sqrt{s_{_{\rm{NN}}}}} = $ 54.4~GeV in the centralities (a) 40-60\%, (b) 60-80\%, (c) 80-100\% and (d) $\ensuremath{\sqrt{s_{_{\rm{NN}}}}} = $ 200~GeV for the 80-100\% centrality range (scaled $\times$0.4), compared to the lowest order EPA-QED predictions (dashed line)~\cite{Wang:2022ihj}. Statistical uncertainties are shown as vertical bars. The systematic uncertainties are shown as blue boxes.}
    \label{fig: momentum transfer}
\end{figure}

\begin{figure}
    \centering
    \includegraphics[width=0.5\textwidth]{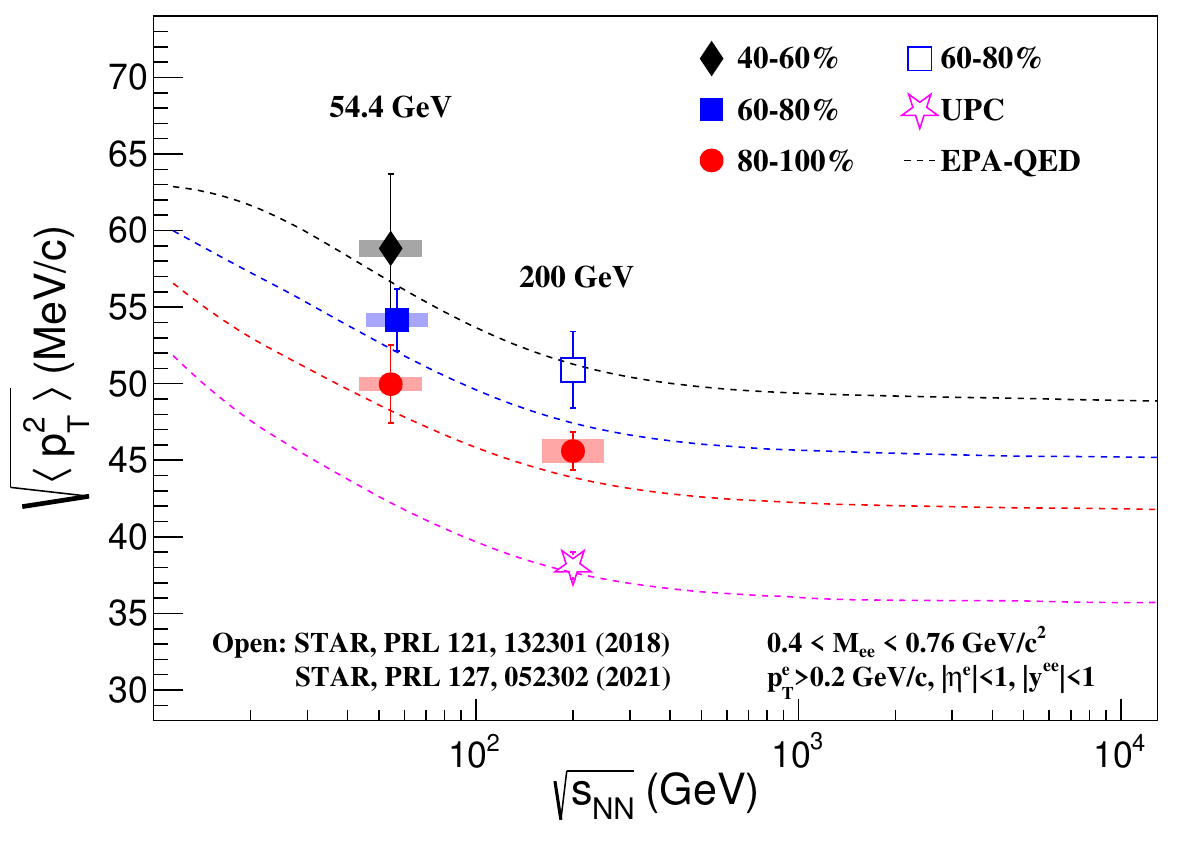}
    \caption{ The $\sqrt{\langle p_{\rm T}^{2} \rangle}$ of $e^{+}e^{-}$ pairs as a function of beam energy compared to the lowest order EPA-QED predictions~\cite{Wang:2022ihj} shown as dashed lines in Au+Au collisions for the centrality intervals of 40-60\%, 60-80\%, 80-100\%, and UPCs. The pair invariant mass region is 0.4-0.76~GeV/$c^{2}$. The solid blue marker is offset slightly in the horizontal direction for clarity. Statistical uncertainties are shown as vertical bars, while systematic uncertainties are shown as boxes. Open markers are extracted from STAR previous published data in Refs.~\cite{STAR:2018ldd,STAR:2019wlg}.}
    \label{fig: mean pT}
\end{figure}

\begin{figure}
    \centering
    \includegraphics[width=0.8\textwidth]{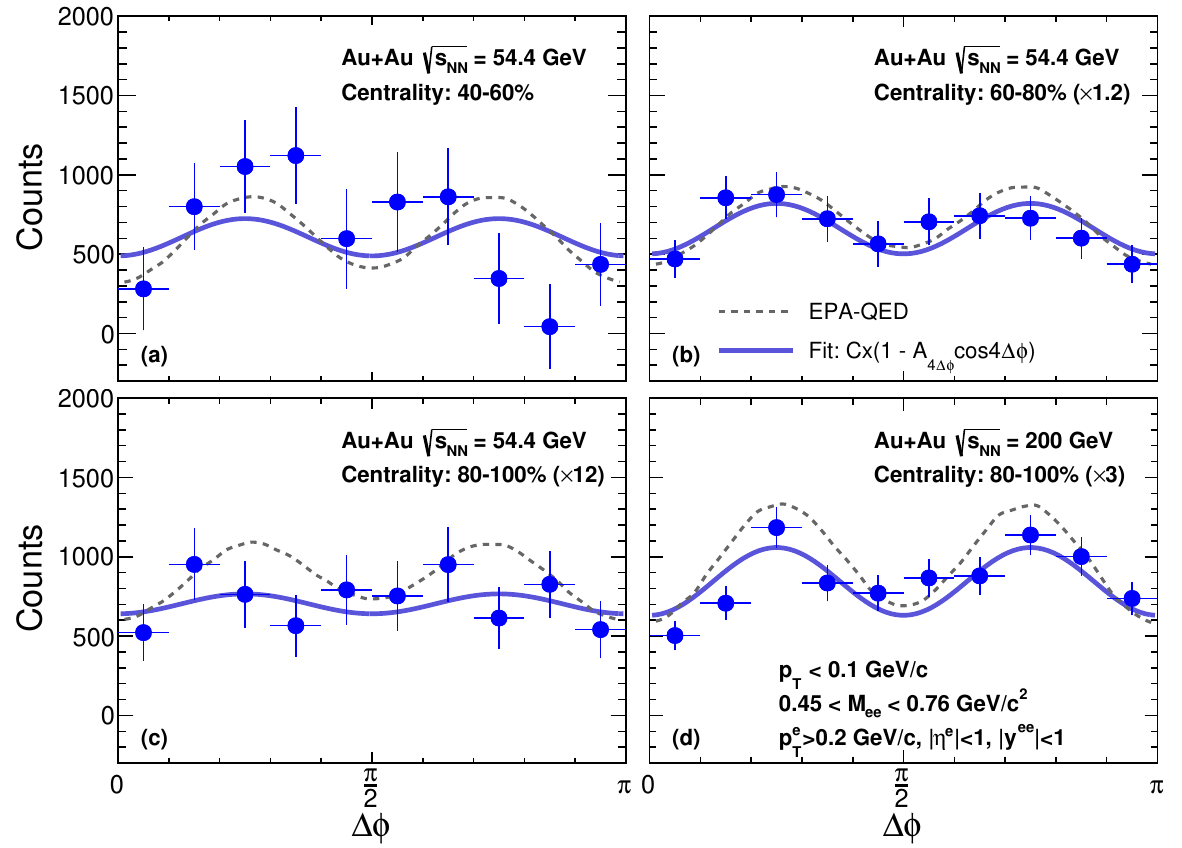}
    \caption{ The $\Delta\phi$ distributions of signal dielectrons in Au+Au collisions at $\ensuremath{\sqrt{s_{_{\rm{NN}}}}} = $ 54.4~GeV in the centralities (a) 40-60\%, (b) 60-80\% (scaled $\times$1.2), (c) 80-100\% (scaled $\times$12) and (d) $\ensuremath{\sqrt{s_{_{\rm{NN}}}}} = $ 200~GeV for the 80-100\% centrality range (scaled $\times$3). The pair invariant mass region is 0.45-0.76~GeV/$c^{2}$ and pair $p_{\rm T} <$~0.1~GeV/$c$. Statistical uncertainties are shown as vertical bars. The full curves are the results of the fits according to Eq.~\eqref{eq: fit function of deltaphi } (solid curves) to be compared with EPA-QED predictions (dashed curves)~\cite{Wang:2022ihj}.}
    \label{fig: delta phi}
\end{figure}

\begin{figure}
    \centering
    \includegraphics[width=0.8\textwidth]{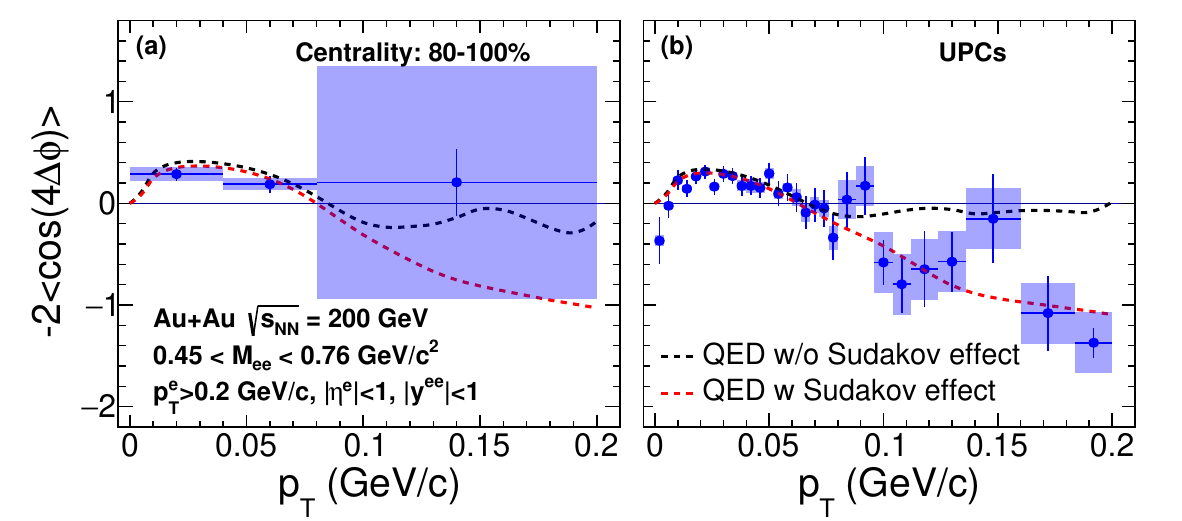}
    \caption{ The amplitude of $\cos(4\Delta\phi)$, ($A_{4\Delta\phi}$), in (a) 80-100\% centrality and (b) UPCs at 200~GeV compared to the different EPA-QED predictions~\cite{Shao:2022stc}. The pair invariant mass region is 0.45-0.76~GeV/$c^{2}$. The black line shows the lowest order QED calculation, while the red line shows the QED calculation with high order effect (Sudakov effect). Statistical uncertainties are shown as vertical bars, while systematic uncertainties are shown as blue boxes.}
    \label{fig: cosphi}
\end{figure}

\begin{figure}
    \centering
    \includegraphics[width=0.8\textwidth]{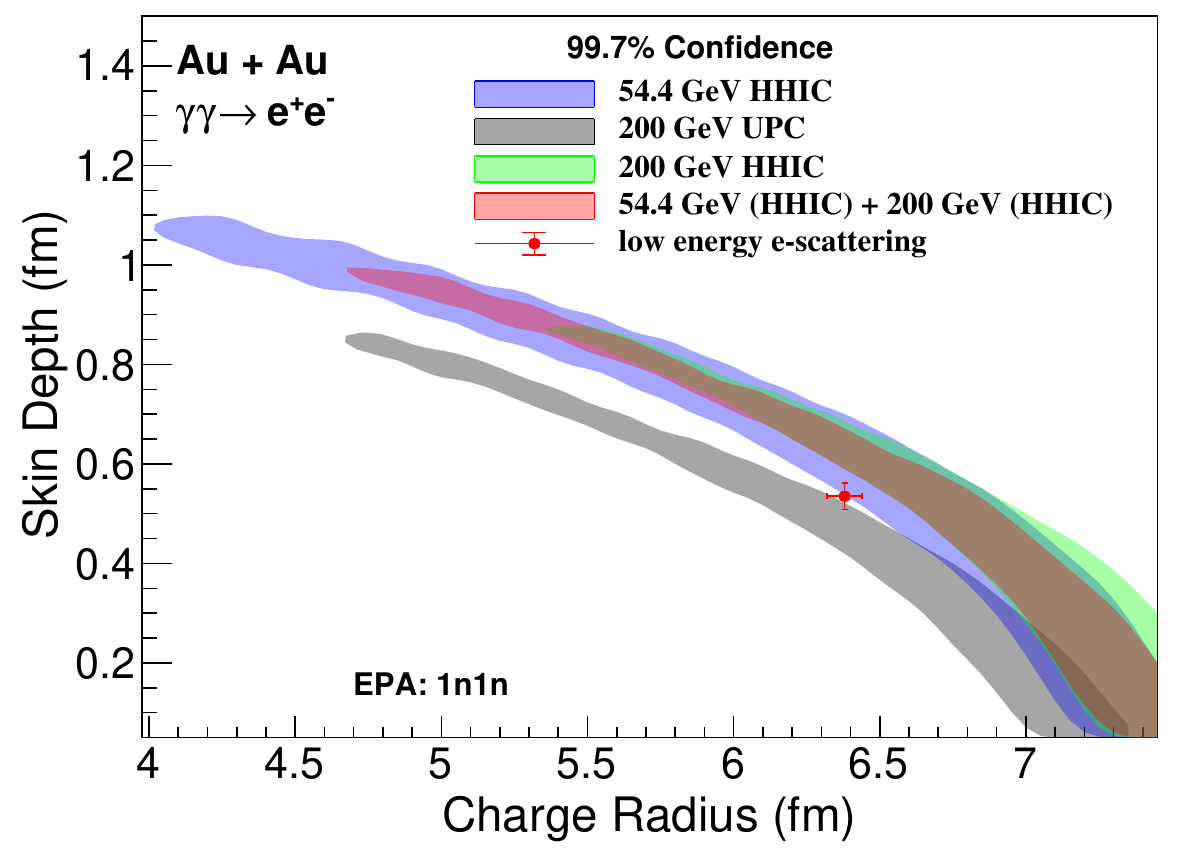}
    \caption{ The constraints on Au nuclear charge distribution extracted by the comparison between STAR measurement of $\gamma\gamma \rightarrow e^{+}e^{-}$ and the corresponding lowest order EPA-QED calculation~\cite{Wang:2022ihj}. HHIC includes 40-60\%, 60-80\% and 80-100\% centralities.}
    \label{fig: contour}
\end{figure}

\subsection{ $p_{\rm T}^{2}$ Distributions}
To further explore the properties of dielectrons produced via the Breit-Wheeler process, the $p_{\rm T}^{2}$ distributions of the signal pairs within STAR acceptance in the invariant mass region of 0.4-0.76~GeV/$c^{2}$ in different centralities at 54.4~GeV and 200~GeV are shown in Fig.~\ref{fig: momentum transfer}. The aforementioned lowest-order numerical EPA-QED calculations are also shown in the plots as dotted lines, which are consistent with the data within uncertainties.
 
Since $\sqrt{\langle p_{\rm T}^{2} \rangle}$ is more sensitive to $p_{\rm T}$ broadening than the distribution of $p_{\rm T}$ itself, we study $\sqrt{\langle p_{\rm T}^{2} \rangle}$ for $e^{+}e^{-}$ pairs as a function of beam energy in the invariant mass region of 0.4-0.76~GeV/$c^{2}$ in different centralities, as shown in Fig.~\ref{fig: mean pT}. Due to statistical limitations, EPA-QED results are used to extrapolate to the unmeasured higher $p_{\rm T}^2$ region to account for the missing contribution. One can see that $\sqrt{\langle p_{\rm T}^{2} \rangle}$ decreases with increasing impact parameter at both beam energies. This results from the impact parameter dependence of the transverse momentum of the initial photon. For high precision results at $\ensuremath{\sqrt{s_{_{\rm{NN}}}}} = $ 200~GeV in UPCs, the consistency between the EPA-QED prediction~\cite{Wang:2022ihj} and our measurement shows that the EPA-QED predictions at $\ensuremath{\sqrt{s_{_{\rm{NN}}}}} = $ 200~GeV can be treated as a baseline. A difference of 4.3~$\sigma$ is found when comparing the data at $\ensuremath{\sqrt{s_{_{\rm{NN}}}}} = $ 54.4~GeV to EPA-QED predictions at $\ensuremath{\sqrt{s_{_{\rm{NN}}}}} = $ 200~GeV, which arises from the energy dependence of $\sqrt{\langle p_{\rm T}^{2} \rangle}$ and possible final-state effects. The observed energy dependence shows that $\sqrt{\langle p_{\rm T}^{2} \rangle}$ decreases with increasing beam energy, which is consistent with EPA-QED predictions. Final-state effects will be discussed in~\Cref{sec:Application: Constrain Charge Distribution}.

\begin{table*}[htbp]
\caption{\label{table: A4phi}The amplitudes ($A_{4\Delta\phi}$) of $\cos(4\Delta\phi)$ and $\chi^{2}/NDF$ of the fits according to Eq.~\eqref{eq: fit function of deltaphi } to be compared with EPA-QED predictions. }
\begin{center}
\begin{tabular}{c|c|c|c|c}
 \hline
 \multirow{2}{*}{} & \multicolumn{2}{c|}{Fit} & \multicolumn{2}{c}{EPA-QED}\\
 \cline{2-5}
  & $A_{4\Delta\phi}$ & $\chi^{2}/NDF$ & $A_{4\Delta\phi}$ & $\chi^{2}/NDF$ \\ [0.5ex] \hline
  54.4 GeV 40-60\%   & 0.19 $\pm$ 0.21 $\pm$ 0.03 & 13.24/8 & 0.40 & 12.40/9 \\[1ex]
  54.4 GeV 60-80\%  & 0.24 $\pm$ 0.09 $\pm$ 0.04 & 4.82/8 & 0.31 & 4.94/9 \\[1ex] 
  54.4 GeV 80-100\% & 0.09 $\pm$ 0.13 $\pm$ 0.01 & 4.98/8 & 0.24 & 13.04/9 \\[1ex] 
  200 GeV 80-100\% & 0.25 $\pm$ 0.06 $\pm$ 0.03 & 12.95/8 & 0.35 & 26.97/9 \\[1ex] \hline
\end{tabular}
\end{center}
\end{table*}

\subsection{$\Delta\phi$ Distributions}

A consequence of the quantum nature of the real photon intrinsic spin and wave function is that the parallel and perpendicular relative polarization angles in photon-photon collisions result in distinct differential cross sections~\cite{Budnev:1975poe,Harland-Lang:2018iur}. It was only recently realized that these effects could be accessed experimentally in ultrarelativistic heavy-ion collisions~\cite{Li:2019yzy} since the transverse momentum of the pair is correlated with the polarization of the photons. For linearly polarized photons, the distinct differential cross sections contains noticeable $\cos(4\Delta\phi)$ and negligible $\cos(2\Delta\phi)$ terms, where $\Delta\phi$ is the azimuthal angle difference between the momentum of the $e^{+}e^{-}$ pair and ($\vec{p_{e^+}}$ - $\vec{p_{e^-}}$) (or ($\vec{p_{e^-}}$ - $\vec{p_{e^+}}$)) in the laboratory frame. Since the photo-induced $e^{+}e^{-}$ pairs are produced almost back-to-back, the $\Delta\phi$ is approximately the azimuthal angle in the laboratory frame between the momentum of the $e^{+}e^{-}$ pair and one of the daughters ($e^{+}$ or $e^{-}$). Recently, STAR has observed $\cos(4\Delta\phi)$ modulations in peripheral and ultra-peripheral Au+Au collisions at 200~GeV, and declared that it is closely related to the phenomenon of vacuum birefringence~\cite{STAR:2019wlg}.

Figure ~\ref{fig: delta phi} shows the $\Delta\phi$ distribution within STAR acceptance for invariant mass region 0.45-0.76~GeV/$c^{2}$ in Au+Au collisions at $\ensuremath{\sqrt{s_{_{\rm{NN}}}}} = $ 54.4~GeV and 200~GeV in different centralities. The fits to a function of the form given as: 

\begin{equation}\label{eq: fit function of deltaphi }
f(\Delta\phi) = C(1 - A_{4\Delta\phi}\cos(4\Delta\phi)),
\end{equation}

\noindent are also shown in Fig.~\ref{fig: delta phi} as solid lines. In Eq.(5), C is a constant, and $A_{4\Delta\phi}$ is the magnitude of $\cos(4\Delta\phi)$ modulation. The observed magnitudes of the $\cos(4\Delta\phi)$ modulations are presented in~\Cref{table: A4phi}, the first and second uncertainties are statistical and systematic uncertainties, respectively. The $\chi^2/NDF$ for the fit of Eq. (5) is also shown in~\Cref{table: A4phi}. In 60-80\% centrality at 54.4~GeV and 80-100\% centrality at 200~GeV, there are statistically significant indications of non-zero $\cos(4\Delta\phi)$ modulation. At 54.4~GeV, due to a large background from the hadronic cocktail in 40-60\% centrality and limited statistics of photon-produced dielectrons in 80-100\% centrality, $\cos(4\Delta\phi)$ modulations are consistent with 0 within uncertainties. Our results are also compared with the lowest-order EPA-QED predictions for the collision of linearly polarized photons, shown in Fig.~\ref{fig: delta phi} as a dashed line. The extracted ($A_{4\Delta\phi}$) and $\chi^{2}/NDF$ values of $\cos(4\Delta\phi)$ are shown in ~\Cref{table: A4phi}.  

The amplitude of $\cos(4\Delta\phi)$, or ($A_{4\Delta\phi}$), is shown as a function of $p_{\rm T}$ in Fig.~\ref{fig: cosphi}(a) for 80-100\% centrality and Fig.~\ref{fig: cosphi}(b) for UPCs at 200~GeV. For UPC events, there is no hadronic or medium-induced background in the selected kinematic range. For more details on the selection and analysis of these events, see Ref.~\cite{STAR:2019wlg}. It can be seen that the $\cos(4\Delta\phi)$ modulation has obvious $p_{\rm T}$ dependence in UPCs. More specifically, below $p_{\rm T}$ $\approx$ 0.08~GeV/$c$ the $\cos(4\Delta\phi)$ modulation amplitude remains positive, then turns increasingly negative for $p_{\rm T}$ above 0.08~GeV/$c$ in UPCs. However, no obvious $p_{\rm T}$ dependence is observed in 80-100\% centrality due to the limited statistics. The figure shows two types of QED calculations~\cite{Shao:2023zge}. They are the lowest order QED shown as the black dashed line, and QED with higher-order effects due to perturbative final state soft photon radiation (Sudakov effect) shown as the red dashed line. The difference between the two QED calculations is small for $p_{\rm T}$ less than about 0.08~GeV/$c$, and STAR statistical precision is not sufficient to distinguish between them so far.  At higher $p_{\rm T}$, however, UPC measurements are consistent with QED when the Sudakov effects are included. Our measurements show the importance of including the radiation of final-state soft photons in UPCs. 

\subsection{Application: Constrain the Au Nuclear Charge Distribution}
\label{sec:Application: Constrain Charge Distribution}

The photon density is related to the energy flux of the electromagnetic fields~\cite{Vidovic:1992ik} $n(\omega) \propto \vec{S} = \frac{1}{\mu_{0}} \vec{E} \times \vec{B}$, where $\mu_{0}$ is vacuum permeability, and $\vec{S}$ is the Poynting Vector. Total and differential cross sections for $\gamma \gamma \rightarrow e^{+}e^{-}$ are related to field strength and spatial distribution. Therefore, assuming the electromagnetic field comes from the charged nucleus, it was proposed that the cross section of $\gamma \gamma \rightarrow e^{+}e^{-}$  can be used to constrain the nuclear charge distribution~\cite{Wang:2022ihj,Brandenburg:2021lnj}. Fig.~\ref{fig: contour} shows the $99.7\%$ ($3\sigma$) confidence level contours for the charge distribution of Au nucleus with different data conditions. These confidence contours result from a $\chi^{2}$-minimization procedure applied to the previous STAR measurements~\cite{STAR:2018ldd,STAR:2019wlg} and these new measurements of the $p_{\rm T}$ and $M_{ee}$ distributions from the $\gamma\gamma \rightarrow e^{+}e^{-}$ process compared to the corresponding lowest-order EPA-QED calculations. For the minimization, the nuclear radius and skin depth are parameterized according to the Woods-Saxon distribution and are assumed to be the same for both electromagnetic and strong interactions. Woods-Saxon distribution is given by:

    \begin{equation}
    \label{equation_charge_density}
    \rho_{A}(r)=\frac{\rho_{0}}{1+\exp[(r-R)/d]},
    \end{equation}

\noindent where $R$ is the nuclear radius and $d$ is the skin depth. The absolute cross section is used to obtain the $\chi^{2}$. The data points in peripheral Au+Au collisions at 54.4~GeV and 200~GeV were used to obtain the blue and green contours, respectively. All available data points in Au+Au peripheral collisions at both 200 GeV~\cite{STAR:2018ldd} and 54.4~GeV were used to obtain the red contour. The data points in ultra-peripheral Au+Au collisions at 200~GeV~\cite{STAR:2019wlg} were used to get the gray contour. In order to incorporate the experimental conditions into the theoretical calculations, the QED calculation in UPCs has included the probability of emitting neutrons from an excited nucleus $1n1n$, where $1n1n$ is defined as two colliding nuclei that each emit a neutron. 

The red marker shown in Fig.~\ref{fig: contour} indicates the result from fits to low energy electron scattering data~\cite{DEVRIES1987495}. The gray contour deviates from blue, green and red contours but is quite close to the red marker, which indicates a potential final-state effect in peripheral hadronic heavy-ion collisions that is not included in the EPA-QED calculations. $e^{+}e^{-}$ pairs produced from photon-photon interactions are mostly back to back, and final-state effects due to trapped magnetic field or Coulomb scattering in the QGP can lead to the observed $p_{\rm T}$ broadening. As Fig.~\ref{fig: mean pT} shows, all non-UPC data points are slightly higher than QED predictions at about the 2.19~$\sigma$ confidence level. For different radius and skin depth, we can get different root-mean-square (RMS) of the radius according to Eq.~\eqref{equation_charge_density}. Then according to the minimum $\chi^{2}$ and the corresponding uncertainty we can get the root-mean-square (RMS) of the charge radius corresponding to the different conditions in Fig.~\ref{fig: contour}, which are listed in \Cref{table: RMS of radius in minimum chi2 plus 1}. These are to be compared to the default value of the nuclear charge radius RMS of $\sqrt{\langle r^2\rangle}=5.33$~fm at $R=6.38$~fm and $d=0.535$~fm~\cite{DEVRIES1987495}. The RMS of the nuclear charge radius extracted by data points at 200~GeV for ultra-peripheral collisions is consistent with the default value, while the result extracted from hadronic heavy-ion collisions at 200~GeV is slightly larger (by about 0.4 fm) than the default value. Although the RMS of the nuclear charge radius at 54.4~GeV HHIC is consistent with the value from low energy electron scattering, the uncertainty is large, and when the 54.4~GeV and 200~GeV HHIC are combined, the large RMS is still favored. These all indicate that a potential final-state effect in hadronic heavy-ion collisions can modify the results of the charge radius extraction and favors an apparent large radius.

\linespread{1.5}
\begin{table*}[htbp]
  \centering
  \caption{\label{table: RMS of radius in minimum chi2 plus 1} The RMS of radius ($\sqrt{\langle r^2\rangle}$) at minimum $\chi^2$ ($\chi^{2}_{min}$) and uncertainties within $\chi^{2}_{min} + 1$.}
  \setlength{\tabcolsep}{5mm}{
\begin{tabular}{|c|c|}
 \hline
 \textbf{ } & \textbf{RMS of Charge Radius (fm)} \\ \hline
 low energy e-scattering experiment & 5.33 $\pm$ 0.05 \\
 200~GeV ( UPC )& 5.39 + 0.16 - 0.21 \\
 54.4~GeV ( HHIC )& 5.39 + 0.16 - 0.30 \\ 
 200~GeV ( HHIC )& 5.72 + 0.07 - 0.04 \\
 54.4~GeV (HHIC) + 200~GeV ( HHIC )& 5.72 + 0.03 - 0.14 \\ \hline
\end{tabular}}
\end{table*}

\section{Conclusions}
\label{sec:conclusions}
We have reported low-$p_{\rm T}$ dielectron measurements in peripheral Au+Au collisions at $\ensuremath{\sqrt{s_{_{\rm{NN}}}}} = $ 54.4~GeV and 200~GeV by the STAR experiment at RHIC. The measured dielectron transverse momentum spectrum shows a significant excess at low-$p_{\rm T}$ ($p_{\rm T} <$~0.15~GeV/$c$) with respect to expected hadronic contributions at both beam energies. The extracted excess yield at low $p_{\rm T}$ as a function of dielectron invariant mass shows a smooth and featureless distribution, which is a consequence of the quantum numbers of the two photons involved in the Breit-Wheeler process. The integrated signal $e^+e^-$ yields in the dielectron invariant mass regions of 0.4-0.76, 0.76-1.2, and 1.2-2.6~GeV/$c^{2}$ show significant energy dependence in different centralities, and the results are consistent with the EPA-QED predictions. $\sqrt{\langle p_{\rm T}^{2} \rangle}$ decreases with increasing impact parameter at both beam energies, while its distribution strongly suggests both an energy dependence and that final-state effects may play a role. The $\Delta\phi$ distribution shows $\cos(4\Delta\phi)$ modulations at $p_{\rm T} <$~0.1~GeV/$c$ with 2.4~$\sigma$ and 3.7~$\sigma$ significance at 54.4~GeV in 60-80\% centrality and 200~GeV in 80-100\% centrality, respectively. Due to statistical limitations, the $\cos(4\Delta\phi)$ modulations in other centralities and energies are consistent with 0 within uncertainties. However, $\cos(2\Delta\phi)$ modulations are found to be consistent with 0 within uncertainty for $p_{\rm T} <$~0.1~GeV/$c$ in all centralities at both beam energies. $-2\langle \cos(4\Delta\phi) \rangle$ shows a clear $p_{\rm T}$ dependence in ultra-peripheral collisions. In the high $p_{\rm T}$ region ($p_{\rm T} >$~0.1~GeV/$c$), the behavior of $-2\langle \cos(4\Delta\phi) \rangle$ as a function of $p_{\rm T}$ indicates that the Sudakov effects need to be included in theoretical calculations. However, this effect does not lead to qualitative conclusions in 80-100\% centrality due to statistical limitations. Finally, we confirmed that measurements of the $\gamma\gamma \rightarrow e^{+}e^{-}$ process can be used to constrain nuclear charge distributions at RHIC energies, though a potential final-state effect in hadronic heavy-ion collisions can modify the extracted nuclear charge radius in a way that favors a larger radius.

\section*{Ackonwledgement}
We thank the RHIC Operations Group and RCF at BNL, the NERSC Center at LBNL, and the Open Science Grid consortium for providing resources and support.  This work was supported in part by the Office of Nuclear Physics within the U.S. DOE Office of Science, the U.S. National Science Foundation, National Natural Science Foundation of China, Chinese Academy of Science, the Ministry of Science and Technology of China and the Chinese Ministry of Education, the Higher Education Sprout Project by Ministry of Education at NCKU, the National Research Foundation of Korea, Czech Science Foundation and Ministry of Education, Youth and Sports of the Czech Republic, Hungarian National Research, Development and Innovation Office, New National Excellency Programme of the Hungarian Ministry of Human Capacities, Department of Atomic Energy and Department of Science and Technology of the Government of India, the National Science Centre and WUT ID-UB of Poland, the Ministry of Science, Education and Sports of the Republic of Croatia, German Bundesministerium f\"ur Bildung, Wissenschaft, Forschung and Technologie (BMBF), Helmholtz Association, Ministry of Education, Culture, Sports, Science, and Technology (MEXT), Japan Society for the Promotion of Science (JSPS) and Agencia Nacional de Investigaci\'on y Desarrollo (ANID) of Chile.

\clearpage

\bibliography{ref}

\end{document}